\documentstyle[12pt,graphicx,amsmath]{article}
\newlength{\pubnumber} 
\catcode`\@=11

\@addtoreset{equation}{section}
\def\section{\@startsection{section}{1}{\z@}{3.5ex plus 1ex minus .2ex}
 {2.3ex plus .2ex}{\large\bf}}
\def\subsection{\@startsection{subsection}{2}{\z@}{2.3ex plus .2ex}
 {2.3ex plus .2ex}{\bf}}
\numberwithin{equation}{section}

\RequirePackage{tabularx}
\RequirePackage{amssymb}

\makeatletter
\renewcommand{\maketag@@@}[1]{\hbox{\m@th\normalsize\normalfont#1}}%
\makeatother

\def\beq{\begin{equation}}
\def\eeq{\end{equation}}
\def\beqn{\begin{eqnarray}}
\def\eeqn{\end{eqnarray}}

\begin{document}

\begin{titlepage}
\samepage{
\setcounter{page}{1}
\rightline{LTH--1028}
\rightline{June 2015}

\vfill
\begin{center}
{\Large \bf Classification\\ \medskip of 
$SU(4) \times SU(2) \times U(1)$ Heterotic-String Models}

\vspace{1cm}
\vfill 

{\large 
Alon E. Faraggi\footnote{alon.faraggi@liv.ac.uk} 
and
Hasan Sonmez\footnote{Hasan.Sonmez@liv.ac.uk}}\\

\vspace{1cm}

{\it Dept.\ of Mathematical Sciences, University of Liverpool, 
Liverpool L69 7ZL, UK\\}

\vspace{.05in}

\vspace{.025in}
\end{center}

\vfill
\begin{abstract}
\noindent
The free fermionic construction of the heterotic string in four
dimensions produced a large space of three generation 
models with the underlying $SO(10)$ embedding of the
Standard Model states. The $SO(10)$ symmetry is broken
to a subgroup directly at the string scale. Over the 
past few years free fermionic models with the Pati--Salam 
and flipped $SU(5)$ subgroups have been classified. 
In this paper we extend this classification program 
to models in which the $SO(10)$ symmetry
is broken at the string level to the
$SU(4)\times SU(2)_L\times U(1)_R$ (SU421) subgroup.
The subspace of free fermionic models that we consider
corresponds to symmetric ${\mathbb{Z}}_2 \times {\mathbb{Z}}_2$ orbifolds. 
We provide a general argument that shows that this 
class of SU421 free fermionic models cannot 
produce viable three generation models. 

\end{abstract}

\smallskip}
\end{titlepage}

\setcounter{footnote}{0}

\newcommand{\cc}[2]{c{#1\atopwithdelims[]#2}}
\newcommand{\nn}{\nonumber}

\def\AEF{A.E. Faraggi}
\def\MODA#1#2#3{{\it Mod.\ Phys.\ Lett.}\/ {\bf A#1} (#2) #3}
\def\IJMP#1#2#3{{\it Int.\ J.\ Mod.\ Phys.}\/ {\bf A#1} (#2) #3}
\def\nuvc#1#2#3{{\it Nuovo Cimento}\/ {\bf #1A} (#2) #3}
\def\RPP#1#2#3{{\it Rept.\ Prog.\ Phys.}\/ {\bf #1} (#2) #3}
\def\EJP#1#2#3{{\it Eur.\ Phys.\ Jour.}\/ {\bf C#1} (#2) #3}
\def\APP#1#2#3{{\it Astropart.\ Phys.}\/ {\bf #1} (#2) #3}
\def\JHEP#1#2#3{{\it JHEP}\/ {\bf #1} (#2) #3}
\def\NPB#1#2#3{{\it Nucl.\ Phys.}\/ {\bf B#1} (#2) #3}
\def\PLB#1#2#3{{\it Phys.\ Lett.}\/ {\bf B#1} (#2) #3}
\def\PRD#1#2#3{{\it Phys.\ Rev.}\/ {\bf D#1} (#2) #3}
\def\PRL#1#2#3{{\it Phys.\ Rev.\ Lett.}\/ {\bf #1} (#2) #3}
\def\PRT#1#2#3{{\it Phys.\ Rep.}\/ {\bf#1} (#2) #3}
\def\etal{{\it et al\/}}

\font\cmss=cmss10 \font\cmsss=cmss10 at 7pt

\hyphenation{space-time-super-sym-met-ric}
\hyphenation{su-per-sym-met-ric non-su-per-sym-met-ric}
\hyphenation{mod-u-lar mod-u-lar--in-var-i-ant}

\setcounter{footnote}{0}
\section{Introduction}

The Standard Model of particle physics accounts successfully 
for all subatomic observational data. The gauge charges
of the Standard Model matter states suggest its embedding
in $SO(10)$ Grand Unified Theory, which is broken to the 
Standard Model at the GUT or string scale. The $SO(10)$
unification picture is further supported by:
the logarithmic evolution of the Standard Model gauge 
parameters; the proton longevity; and the suppression of 
left--handed neutrino masses. The heterotic--string \cite{gross}
produces chiral $SO(10)$ representations in its perturbative
spectrum, and is therefore the one suited to explore 
the $SO(10)$ GUTs structure underlying the Standard Model.
Phenomenological studies of the heterotic--string 
have been pursued since the mid--eighties \cite{candelas}, using 
a variety of world--sheet \cite{fff, gepner, bert} 
and target space techniques \cite{cy, orbifolds}. 

The free fermionic construction of the heterotic--string in four 
dimensions produced a rich space of phenomenological three generation
models. These models admit the underlying $SO(10)$ GUT embedding 
of the Standard Model spectrum. However, the $SO(10)$ symmetry is 
broken directly at the string level. The early 
studies of these models consisted of isolated 
examples that shared an underlying NAHE--base structure \cite{nahe}. 
Examples in which the $SO(10)$ symmetry is broken to the: 
flipped $SU(5)$ (FSU5) \cite{revamp};
$SO(6)\times SO(4)$ Heterotic String Pati--Salam Models (HSPSM) \cite{alr};
$SU(3)\times SU(2)\times U(1)^2$ Standard--like Models (SLM) \cite{slm};
$SU(3)\times SU(2)^2\times U(1)$ left--right symmetric (LRS) \cite{lrs}; 
and $SU(4)\times SU(2)\times U(1)$ (SU421) \cite{su421};
subgroups were studied. 
Among those the FSU5; SLM; HSPSM; LRS cases produced quasi--realistic 
three generation models, whereas the SU421 case did not produce any 
viable three generation model. The advantage of the SU421 models 
compared to the FSU5 and HSPSM is that they admit both the doublet--triplet, 
as well as the doublet--doublet spitting mechanism \cite{su421}. 
We also note the recent interest in SU421 models from purely phenomenological 
considerations \cite{wise}. 

The phenomenological free fermionic heterotic--string models 
are ${\mathbb{Z}}_2 \times {\mathbb{Z}}_2$ orbifolds that are constructed
at enhanced symmetry points in the moduli space
\cite{Z2Z2Faraggi1994,Z2Z2Kounnas1997}.
Many of the phenomenological properties of the models 
are rooted in their underlying ${\mathbb{Z}}_2 \times {\mathbb{Z}}_2$ structure 
\cite{recentreview}.
In recent years
systematic methods for the classification of symmetric 
${\mathbb{Z}}_2 \times {\mathbb{Z}}_2$ free fermionic orbifolds were developed 
in \cite{typeIIclassi} for type II superstrings
and in refs. \cite{fknr,fkr} for 
symmetric ${\mathbb{Z}}_2 \times {\mathbb{Z}}_2$ 
heterotic--string orbifolds with $SO(10)$ GUT symmetry. 
The classification was extended in refs. \cite{acfkr, cfr, rizos} 
and \cite{frs} 
to string vacua in which the $SO(10)$ symmetry is
broken to the $SO(6)\times SO(4)$ Pati--Salam 
and to the flipped $SU(5)$ 
subgroups, respectively. 
The Pati--Salam class of free fermionic vacua
produced examples of three generation exophobic 
models in which exotic fractionally charged states
only appear in the massive string spectrum \cite{acfkr,cfr},
whereas the flipped $SU(5)$ 
class of models did not produce exophobic models 
with an odd number of generations \cite{frs}. 

In this paper we discuss the classification for the 
class of SU421 heterotic--string models. We provide
a general argument that breaking the $SO(10)$ symmetry
to this subgroup cannot produce three chiral generations 
in the prevalent free fermionic construction which is 
based on symmetric
${\mathbb{Z}}_2 \times {\mathbb{Z}}_2$ toroidal compactification 
with a ${\mathbb{Z}}_2 \times {\mathbb{Z}}_4$ fermionic boundary conditions that
break the $SO(10)$ symmetry to $SU(4)\times SU(2)\times U(1)$.

\section{$SU(4) \times SU(2) \times U(1)$ Phenomenology}\label{analysis}

The field theory content of the $N=1$ supersymmetric 
$SU(4)_C\times SU(2)_L\times U(1)_R$ 
model\footnote{we note that $U(1)_R$ as defined here
is equal to $1/2\, U(1)_L$ as defined in ref. \cite{slm}. }
was discussed in 
ref. \cite{su421}.
The SU421 class of heterotic--string 
models differs from the HSPSM models in the breaking 
of $SU(2)_R\rightarrow U(1)_R$ directly at the string
level. Similar to the HSPSM, the SU421 heterotic--string models
admit the $SO(10)$ embedding and the chiral states
are obtained from the spinorial {\bf 16} representations of $SO(10)$ which 
decomposes in the following way: 
\beqn
F_L^{i} &=& (      4 ,2,~~0)     ~~=~
	    (3 ,2, ~~{1\over3},~~~0) + (1,2, -{1},~~~0)
       ~~=~  {u\choose d}^i+{\nu\choose e}^i,\label{SU421fl}\\
U_R^{i} &=& ({\overline 4},1,-{1\over2})    ~=~
	    ({\overline 3},1,-{1\over3},-{1\over2}) + (1,1,+{1},-{1\over2})
       ~~=~{u^{c}}^i+{N^{c}}^i,\label{SU421ur}\\
D_R^{i} &=& ({\overline 4},1,+{1\over2})    ~=~
	    ({\overline 3},1,-{1\over3},+{1\over2}) + (1,2,+{1},+{1\over2})
       ~~=~{d^{c}}^i+{e^{c}}^i. 
\label{SU421dr}
\eeqn
The first and second equalities show the decomposition under 
$SU(4)_C\times SU(2)_L\times U(1)_R$ and
$SU(3)_C\times SU(2)_L\times U(1)_{B-L}\times U(1)_R$, respectively. 
The electroweak $U(1)_Y$ current is given by
\beq
U(1)_Y={1\over2}U(1)_{B-L}+U(1)_R.
\label{ewu1current}
\eeq
From eq. (\ref{SU421fl}) we note that 
$F_L$ produces the quarks and leptons weak doublets, 
and that $U_R$ and $D_R$ produces the right--handed 
weak singlets. The two Higgs multiplets of the Minimal
Supersymmetric Standard Model, $h^u$ and $h^d$, are
given by, 
\beqn
h^d     &=& (      1 ,2,-{1\over2}),  \\
h^u     &=& (      1 ,2,+{1\over2}).
\label{SU421mssmhigss}
\eeqn
The heavy Higgs states that are responsible for
breaking 
$SU(4)_C\times U(1)_{R}$ gauge symmetry to the 
Standard Model groups $SU(3)\times U(1)_Y$
are given by the fields
\beqn
     {H} &=& ({\overline 4},1,-{1\over 2})\\
{\overline H} &=& ({     4},1,+{1\over 2})
\label{SU421Higgs}
\eeqn
The SU421 heterotic--string models may also contain states that transform as
$$(6,1,0)= ({3},1,{1\over3},0)+ ({\overline 3},1,-{1\over3},0).$$
These multiplets  
arise from the vectorial {\bf 10} representation of $SO(10)$.
These coloured states generate proton decay from dimension five operators,
and therefore must be sufficiently heavy to be in agreement with
the proton lifetime limits. An important benefit of the SU421 
symmetry breaking pattern is that these colour triplets may 
be projected out by the Generalised GSO (GGSO) projections \cite{ps}, 
and need not be present in the low energy spectrum. 
The string doublet--triplet mechanism works 
in all models that include the symmetry 
breaking pattern $SO(10)\rightarrow SO(6)\times SO(4)$.
The HSPSM heavy Higgs states, which break
$SU(4)\times SU(2)_R\rightarrow SU(3)_C\times U(1)_{Y}$, 
contain colour triplets with the charges of the states 
in (\ref{SU421dr}) that may give rise to dimension five
proton decay mediating operators. In the HSPSM the superpotential
terms $\lambda_2HHD+\lambda_3 {\bar H}{\bar H}{\bar D}$
couples the colour
triplets from the vectorial representation $(6,1,1)$ to 
the colour triplets arising from the heavy Higgs field.
The GUT scale VEVs of the heavy Higgs fields 
$H$ and ${\bar H}$ are used to give heavy mass
to the Higgs colour triplets. 
However, the heavy Higgs representations in the 
SU421 heterotic--string models, eq. (\ref{SU421Higgs}),
do not contain the states with the charges of
eq. (\ref{SU421dr}). Consequently, the stringy
doublet--triplet splitting mechanism works
only in models in which the $SO(10)$ symmetry is broken 
to $SU(3)_C\times SU(2)_L\times U(1)^2$,
$SU(4)_C\times SU(2)_L\times U(1)_R$, or
$SU(3)_C\times SU(2)_L\times SU(2)_R \times U(1)_{B-L}$.

Another important advantage of the SU421 class of 
models versus the PS and LRS models is with respect
to the light Higgs representations. In the LRS and PS 
models, the light Higgs states exist in bi--doublet
representations and couple simultaneously to the 
up-- and down--type quarks, which may  give 
rise to Flavor Changing Neutral Currents (FCNC) 
at an unacceptable rate \cite{lrsfcnc}.
This introduces a bi--doublet splitting problem. 
The solutions that have been proposed in the literature
\cite{bidoub} use a $SU(2)_L$ triplet representation
that are not present in string models in which the 
gauge symmetry is realised as a level one Kac--Moody algebra. 
On the other hand, in SU421 models $SU(2)_R$ is broken 
at the string level and consequently the Higgs bi--doublet
is split at the string level. 

The solutions to the doublet--doublet as well as the
doublet--triplet splitting problems are the two appealing 
properties offered by the SU421 free fermionic 
heterotic--string models. However, as we argue in the
next section the free fermionic ${\mathbb{Z}}_2 \times {\mathbb{Z}}_2$ 
orbifold models, with additional ${\mathbb{Z}}_2 \times {\mathbb{Z}}_4$
basis vectors that are used to break the 
$SO(10)$ symmetry to $SU(4)_C\times SU(2)_L\times U(1)_R$,
cannot in fact produce three complete chiral generations 
and therefore, like the NAHE--based free fermionic 
models \cite{su421}, these models do not produce 
viable SU421 string models.  

\subsection{The $SU(4) \times SU(2) \times U(1)$ Free Fermionic Construction}

The string vacuum in the free fermionic formulation \cite{fff}
is defined in terms of a set of boundary condition basis 
vectors and the Generalised GSO projection coefficients,
which span the one--loop partition function. 
The basis vectors generate a finite additive group
$\Xi=\sum_k{{n_k}{b_k}}$
where $n_k=0,\cdots,{{N_{z_k}}-1}$.
The physical states in the Hilbert space of a sector
$\alpha\in{\Xi}$ are obtained by acting on the vacuum
with fermionic and bosonic oscillators and by applying 
the GGSO projections. Each fermionic complex oscillator
acting on the vacuum is counted by a fermion number
operator as $F_\alpha(f)=1$ and $\alpha(f^*)=-1$.
For periodic complex
fermions with $\alpha(f)=1$, 
the vacuum is in a doubly degenerate spinorial representation
${\vert \pm\rangle}$, 
annihilated by the zero modes $f_0$ and
${{f_0}^*}$ and with fermion numbers $F(f)=0,-1$, respectively.
The $U(1)$ charges $Q(f)$ of the unbroken Cartan generators of
the right--moving gauge group are given in terms of the boundary 
conditions and fermion numbers of the complex right--moving 
world--sheet fermions by
\begin{equation}
{Q(f) = {1\over 2}\alpha(f) + F(f)}. 
\label{u1charges}
\end{equation}

In the light--cone gauge, the free fermionic heterotic--string
models in four
dimensions require $20$ and $44$, left--moving and 
right--moving real world--sheet fermions respectively, 
to cancel the conformal anomaly. 
In the usual notation these are denoted as: 
$\psi^\mu, \chi^{1,\dots,6},y^{1,\dots,6}, \omega^{1,\dots,6}$ and 
$\overline{y}^{1,\dots,6},\overline{\omega}^{1,\dots,6}$,
$\overline{\psi}^{1,\dots,5}$, $\overline{\eta}^{1,2,3}$, 
$\overline{\phi}^{1,\dots,8}$. 

\subsection{The $SU(4) \times SU(2) \times U(1)$ Gauge Group}

In the following we set up the necessary ingredients for the
classification of the SU421 free fermionic heterotic--string models. 
The analysis is along similar lines to the one performed 
in the classification of the $SO(10)$ \cite{fknr}; 
heterotic--string Pati--Salam models \cite{acfkr}; and 
flipped $SU(5)$ models \cite{frs}. The novelty compared to these
cases is that the SU421 models employ two basis vectors 
that break the $SO(10)$ symmetry, whereas the HSPSM and FSU5 models 
use only one. However, we argue below that this class of heterotic--string
vacua cannot in fact produce phenomenologically viable models. 
The basis vectors that generate our $SU(4) \times SU(2) \times U(1)$ 
heterotic--string models are given by the following 14 basis vectors
\begin{eqnarray} \label{421}
v_1={\bf1}&=&\{\psi^\mu,\
\chi^{1,\dots,6},y^{1,\dots,6}, \omega^{1,\dots,6}|\overline{y}^{1,\dots,6},
\overline{\omega}^{1,\dots,6},
\overline{\eta}^{1,2,3},
\overline{\psi}^{1,\dots,5},\overline{\phi}^{1,\dots,8}\},\nonumber\\
v_2=S&=&\{{\psi^\mu},\chi^{1,\dots,6}\},\nonumber\\
v_{2+i}={e_i}&=&\{y^{i},\omega^{i}|\overline{y}^i,\overline{\omega}^i\}, \
i=1,\dots,6,\nonumber\\
v_{9}={b_1}&=&\{\chi^{34},\chi^{56},y^{34},y^{56}|\overline{y}^{34},
\overline{y}^{56},\overline{\eta}^1,\overline{\psi}^{1,\dots,5}\},\label{basis}\\
v_{10}={b_2}&=&\{\chi^{12},\chi^{56},y^{12},y^{56}|\overline{y}^{12},
\overline{y}^{56},\overline{\eta}^2,\overline{\psi}^{1,\dots,5}\},\nonumber\\
v_{11}=z_1&=&\{\overline{\phi}^{1,\dots,4}\},\nonumber\\
v_{12}=z_2&=&\{\overline{\phi}^{5,\dots,8}\},\nonumber\\
v_{13}=\alpha&=&\{\overline{\psi}^{4,5},\overline{\phi}^{1,2}\},\nonumber\\
v_{14}=\beta&=&\{\overline{\psi}^{4,5}=\textstyle\frac{1}{2},
\overline{\phi}^{1,...,6}=\textstyle\frac{1}{2}\}.
\nonumber
\end{eqnarray}
The basis vector {\bf1} generates models with $SO(44)$ gauge 
group from the Neveu--Schwarz sector. The vector $S$
produces ${N} = 4$ space--time supersymmetry. 
The vectors $e_{1}$,$\dots$,$e_{6}$ break the $SO(44)$ gauge group to 
$SO(32) \times U(1)^6$ and preserve the ${N = 4}$ space--time supersymmetry. 
The $e_i$ basis vectors correspond to all the possible symmetric shifts of 
the six internal bosonic coordinates. 
The basis vectors $b_1$ and $b_2$  correspond to 
${\mathbb{Z}}_2 \times {\mathbb{Z}}_2$ 
orbifold twists and break ${N} = 4$ space--time supersymmetry to
$N=1$. Additionally, they reduce the rank of gauge group by breaking 
the $U(1)^6$ symmetry. Combined with the projections of the basis
vectors $z_{1}$ and $z_{2}$ the $SO(32)$ gauge group is reduced to
$SO(10) \times U(1)^3 \times SO(8)_1\times SO(8)_2$, 
where $SO(10) \times U(1)^3$ and $SO(8)_1\times SO(8)_2$ 
correspond to the observable and hidden gauge groups, respectively. 
The combined projection of the basis vectors $\alpha$ and $\beta$ 
breaks the $SO(10)$ GUT symmetry to $SU(4) \times SU(2) \times U(1)$, 
where $\alpha$ is identical to the basis vector used in the classification of
the Pati--Salam models, and hence breaks the $SO(10)$ symmetry to 
$SO(6)\times SO(4)$ and finally using the $\beta$ basis vector 
with fractional boundary conditions reduces the $SO(10)$
gauge symmetry to $SU(4) \times SU(2) \times U(1)$. 

\subsection{The String Spectrum}\label{analysisspec}

The space--time vector bosons that are obtained from the 
Neveu--Schwarz (NS) sector and that survive the GGSO projections, 
defined by the basis vectors in (\ref{basis}) 
generate the observable and hidden gauge groups 
given by:  
\begin{eqnarray}
{\rm Observable} &: &~~~~SU(4) \times SU(2)_L 
                         \times U(1)_R \times{U(1)}^3 \nonumber\\
{\rm Hidden}     &: &~~~~SU(2)_A \times U(1)_A \times SU(2)_B 
                         \times U(1)_B \times SU(2)_C \times U(1)_C 
                         \times SO(4)_2 \nonumber
\end{eqnarray}
The string states arising in other sectors transform under these 
gauge group factors. 
Additional space--time vector bosons that enhance the NS observable 
and/or hidden gauge groups may arise from additional sectors. 
In order to preserve the above gauge groups, all these additional
space--time vector bosons 
need to be projected out. These additional 
space--time vector bosons arise from the following 36 sectors
\small
\begin{equation}
\mathbf{G}_{Enh} =
\left\{ \begin{array}{ccccc}
\,\,\,\, z_1          ,&
\,\,\,\, z_1 + \beta          ,&
\,\,\,\, z_1 + 2\beta    ,\\
\,\,\,\,\, z_1 + \alpha       ,&
\,\,\,\,\, z_1 + \alpha + \beta ,&
\,\,\, z_1 + \alpha + 2\beta ,\\
\,\, z_2       ,&
\,\,\,\,\,\,\,\, z_2 + \beta ,&
\,\,\,\,\,\, z_2 + 2\beta ,\\
\,\,\,\,\, z_2 + \alpha       ,&
\,\,\,\,\, z_2 + \alpha + \beta ,&
\,\,\, z_2 + \alpha + 2\beta ,\\
\,\, z_1 + z_2       ,&
\,\,\,\,\,\,\,\, z_1 + z_2 + \beta ,&
\,\,\,\,\,\, z_1 + z_2 + 2\beta ,\\
\,\,\,\,\, z_1 + z_2 + \alpha       ,&
\,\,\,\,\, z_1 + z_2 + \alpha + \beta ,&
\,\,\, z_1 + z_2 + \alpha + 2\beta ,\\
\,\, \beta       ,&
\,\,\,\,\,\,\,\, 2\beta ,&
\,\,\, \alpha,\\
\,\,\,\,\, \alpha + \beta       ,&
\,\,\,\,\, \alpha + 2\beta ,&
\,\, x,\\
\,\,\,\,\, z_1 + x + \beta       ,&
\,\,\,\,\, z_1 + x + 2\beta ,&
\,\, z_1 + x + \alpha,\\
\,\,\,\,\, z_1 + x + \alpha + \beta       ,&
\,\,\, z_2 + x + \beta ,&
\,\,\,\,\, z_2 + x + \alpha + \beta       ,\\
\,\,\, z_1 + z_2 + x + \beta ,&
\,\, z_1 + z_2 + x + 2\beta,&
\,\,\,\,\, z_1 + z_2 + x + \alpha + \beta ,\\
\,\, x + \beta,&
\,\,\,\,\, x + \alpha ,&
\,\,\, x + \alpha + \beta ,\\
\end{array} \right\}, \label{ggsectors1}
\end{equation}
\\
\normalsize
where $x = 1 + S + \textstyle\sum_{i = 1}^{6} e_i + z_1 + z_2$.

\subsection{The Matter Content}\label{analysis2}
The observable matter states in heterotic--string vacuum with $(2,2)$
world--sheet supersymmetry is embedded in the 
$\bf{27}$ representation of $E_6$. 
In the free fermionic construction that we adopt here, and 
using the basis vectors in (\ref{421}), the $E_6$ is first
broken to the $SO(10)\times U(1)$ symmetry. 
Therefore, the $\bf{27}$ of $E_6$ decomposes in the following way
\begin{eqnarray}
\textbf{27} &= & \textbf{16} + \textbf{10} + \textbf{1}.
\end{eqnarray}
Where the $\textbf{16}$ transforms under the spinorial representation of 
$SO(10)$ and \textbf{10} transforms in the vectorial representation of the $SO(10)$, 
and similarly for 
$\bf{\overline{27}}$. 
The following 48 sectors produce states that give the spinorial 
$\bf{16}$ or $\bf{\overline{16}}$ of $SO(10)$
\begin{eqnarray} \label{obspin}
B_{pqrs}^{(1)}&=& S + {b_1 + p e_3+ q e_4 + r e_5 + s e_6} \nonumber\\
&=&\{\psi^\mu,\chi^{12},(1-p)y^{3}\overline{y}^3,p\omega^{3}\overline{\omega}^3,
(1-q)y^{4}\overline{y}^4,q\omega^{4}\overline{\omega}^4, \nonumber\\
& & ~~~(1-r)y^{5}\overline{y}^5,r\omega^{5}\overline{\omega}^5,
(1-s)y^{6}\overline{y}^6,s\omega^{6}\overline{\omega}^6,
\overline{\eta}^1,\overline{\psi}^{1,...,5}\},
\\
B_{pqrs}^{(2)}&=& S + {b_2 + p e_1+ q e_2 + r e_5 + s e_6},
\label{twochiralspinorials}
\nonumber\\
B_{pqrs}^{(3)}&=& S + {b_3 + p e_1+ q e_2 + r e_3 + s e_4}, \nonumber
\end{eqnarray}
where $p,q,r,s=0,1$ and $b_3=b_1+b_2+x$. 
In order to distinguish between the spinorial $\bf{16}$ and $\bf{\overline{16}}$ in 
the states given above, the following chirality operators are used

\begin{eqnarray}\label{so10operators}
X_{pqrs}^{(1)_{SO(10)}} & = &
C\binom{B^{(1)}_{pqrs}}{b_{2} + (1-r)e_{5} + (1-s)e_{6}},\nonumber\\
X_{pqrs}^{(2)_{SO(10)}} & = &
C\binom{B^{(2)}_{pqrs}}{b_{1} + (1-r)e_{5} + (1-s)e_{6}},\\
X_{pqrs}^{(3)_{SO(10)}} & = &
C\binom{B^{(3)}_{pqrs}}{b_{1} + (1-r)e_{3} + (1-s)e_{4}}.\nonumber
\end{eqnarray}
Where $X_{pqrs}^{(1,2,3)_{SO(10)}} = 1$ implies the states corresponds to the 
$\bf{16}$ of $SO(10)$ and $X_{pqrs}^{(i)_{SO(10)}} = -1$ to the 
$\bf{\overline{16}}$ of $SO(10)$. 
Moreover, we note that the states here can be projected in or out depending on 
the GGSO projections of the basis vectors $e_1,....,e_6$, $z_1$ and $z_2$. 
Therefore, we define below a projector $P$, such that $P=1$ implies the state is projected in 
and $P=0$ implies the state is projected out. The projector $P$ is given by

\footnotesize
\begin{align}\label{matrixequations}
P_{pqrs}^{(1)} &= \frac{1}{16} 
\left( 1-C \binom {e_1} {B_{pqrs}^{(1)}}\right) . 
\left( 1-C \binom {e_2} {B_{pqrs}^{(1)}}\right) . 
\left( 1-C \binom {z_1} {B_{pqrs}^{(1)}}\right) . 
\left( 1-C \binom {z_2} {B_{pqrs}^{(1)}}\right),\nonumber\\
P_{pqrs}^{(2)} &= \frac{1}{16} 
\left( 1-C \binom {e_3} {B_{pqrs}^{(2)}}\right) . 
\left( 1-C \binom {e_4} {B_{pqrs}^{(2)}}\right) . 
\left( 1-C \binom {z_1} {B_{pqrs}^{(2)}}\right) . 
\left( 1-C \binom {z_2} {B_{pqrs}^{(2)}}\right),\\
P_{pqrs}^{(3)} &= \frac{1}{16} 
\left( 1-C \binom {e_5} {B_{pqrs}^{(3)}}\right) . 
\left( 1-C \binom {e_6} {B_{pqrs}^{(3)}}\right) . 
\left( 1-C \binom {z_1} {B_{pqrs}^{(3)}}\right) . 
\left( 1-C \binom {z_2} {B_{pqrs}^{(3)}}\right).\nonumber
\end{align}
\normalsize
These projectors above can in fact be expressed as matrix equations given by  

\begin{align}
\begin{pmatrix} (e_1|e_3)&(e_1|e_4)&(e_1|e_5)&(e_1|e_6)\\
(e_2|e_3)&(e_2|e_4)&(e_2|e_5)&(e_2|e_6)\\
(z_1|e_3)&(z_1|e_4)&(z_1|e_5)&(z_1|e_6)\\
(z_2|e_3)&(z_2|e_4)&(z_2|e_5)&(z_2|e_6) \end{pmatrix}
\begin{pmatrix} p\\q\\r\\s\end{pmatrix} &=
\begin{pmatrix} (e_1|b_1)\\
(e_2|b_1)\\
(z_1|b_1)\\
(z_2|b_1)
\end{pmatrix},\nonumber
\\[0.3cm]
\begin{pmatrix} (e_3|e_1)&(e_3|e_2)&(e_3|e_5)&(e_3|e_6)\\
(e_4|e_1)&(e_4|e_2)&(e_4|e_5)&(e_4|e_6)\\
(z_1|e_1)&(z_1|e_2)&(z_1|e_5)&(z_1|e_6)\\
(z_2|e_1)&(z_2|e_2)&(z_2|e_5)&(z_2|e_6) \end{pmatrix}
\begin{pmatrix} p\\q\\r\\s\end{pmatrix} &=
\begin{pmatrix} (e_3|b_2)\\
(e_4|b_2)\\
(z_1|b_2)\\
(z_2|b_2)
\end{pmatrix},
\\[0.3cm]
\begin{pmatrix} (e_5|e_1)&(e_5|e_2)&(e_5|e_3)&(e_5|e_4)\\
(e_6|e_1)&(e_6|e_2)&(e_6|e_3)&(e_6|e_4)\\
(z_1|e_1)&(z_1|e_2)&(z_1|e_3)&(z_1|e_4)\\
(z_2|e_1)&(z_2|e_2)&(z_2|e_3)&(z_2|e_4) \end{pmatrix}
\begin{pmatrix} p\\q\\r\\s\end{pmatrix} &=
\begin{pmatrix} (e_5|b_3)\\
(e_6|b_3)\\
(z_1|b_3)\\
(z_2|b_3)
\end{pmatrix}.\nonumber
\end{align}
Writing the projectors as matrix equations given above entails
solving systems of linear equations. 
These algebraic equations can be solved using a computerised 
code, which can be used to scan a vast space of models.

Similar to the spinorial representations singlet and
vectorial $\bf{10}$ representations of $SO(10)$ 
are obtained from the following 48 sectors
\begin{eqnarray}\label{lighthiggssectors}
B_{pqrs}^{(4)}&=& B_{pqrs}^{(1)} + x
\nonumber\\
&=&\{\psi^\mu,\chi^{12},(1-p)y^{3}\overline{y}^3,p\omega^{3}\overline{\omega}^3,
(1-q)y^{4}\overline{y}^4,q\omega^{4}\overline{\omega}^4, \nonumber\\
& & ~~~~~~~~~(1-r)y^{5}\overline{y}^5,r\omega^{5}\overline{\omega}^5,
(1-s)y^{6}\overline{y}^6,s\omega^{6}\overline{\omega}^6,\overline{\eta}^{2,3} \},
\label{nonchiralvectorials}\\
B_{pqrs}^{(5,6)}&=& B_{pqrs}^{(2,3)} + x. \nonumber
\end{eqnarray}
Massless states that arise in these sectors are obtained by acting on
the vacuum with a NS oscillator. The type of states therefore depend 
on the type of oscillator, and may correspond to $SO(10)$ singlets 
or vectorial $\bf{10}$ representation of
$SO(10)$, which is needed for electroweak symmetry breaking. 
The different type of $SO(10)$ singlets arising from eq.
(\ref{nonchiralvectorials}) are 
\begin{itemize}
\item $\{\overline\eta^{i}\}|R \rangle_{pqrs}^{(4,5,6)}$ or
$\{\overline\eta^{*i}\}|R\rangle_{pqrs}^{(4,5,6)}$, $i = 1,2,3$,
where $|R\rangle_{pqrs}^{(4,5,6)}$ is the degenerated Ramond vacuum of the
$B_{pqrs}^{(4,5,6)}$ sector.
These states transform as a vector--like representations under the $U(1)_i$'s.

\item $\{\overline\phi^{1,2}\}|R\rangle_{pqrs}^{(4,5,6)}$ or
$\{\overline\phi^{*1,2}\}|R\rangle_{pqrs}^{(4,5,6)}$.
These states transform as a vector--like representations 
of $SU(2)_A \times U(1)_A$.

\item $\{\overline\phi^{3,4}\}|R\rangle_{pqrs}^{(4,5,6)}$ or
$\{\overline\phi^{*3,4}\}|R\rangle_{pqrs}^{(4,5,6)}$.
These states transform as a vector--like representations 
of $SU(2)_B \times U(1)_B$.

\item $\{\overline\phi^{5,6}\}|R\rangle_{pqrs}^{(4,5,6)}$ or
$\{\overline\phi^{*5,6}\}|R\rangle_{pqrs}^{(4,5,6)}$.
These states transform as a vector--like representations 
of $SU(2)_C \times U(1)_C$.

\item $\{\overline\phi^{7,8}\}|R\rangle_{pqrs}^{(4,5,6)}$ or
$\{\overline\phi^{*7,8}\}|R\rangle_{pqrs}^{(4,5,6)}$.
These states transform as a vector--like representations 
of $SO(4)$.
\end{itemize}
Similarly, for the matrix equations given above in
eq. (\ref{matrixequations}), we can write algebraic equations
for the sectors in eq. (\ref{lighthiggssectors}) given as follows:

\begin{align}
\begin{pmatrix} (e_1|e_3)&(e_1|e_4)&(e_1|e_5)&(e_1|e_6)\\
(e_2|e_3)&(e_2|e_4)&(e_2|e_5)&(e_2|e_6)\\
(z_1|e_3)&(z_1|e_4)&(z_1|e_5)&(z_1|e_6)\\
(z_2|e_3)&(z_2|e_4)&(z_2|e_5)&(z_2|e_6) \end{pmatrix}
\begin{pmatrix} p\\q\\r\\s\end{pmatrix} &=
\begin{pmatrix} (e_1|b_1 + x)\\
(e_2|b_1 + x)\\
(z_1|b_1 + x)\\
(z_2|b_1 + x)
\end{pmatrix},\nonumber
\\[0.3cm]
\begin{pmatrix} (e_3|e_1)&(e_3|e_2)&(e_3|e_5)&(e_3|e_6)\\
(e_4|e_1)&(e_4|e_2)&(e_4|e_5)&(e_4|e_6)\\
(z_1|e_1)&(z_1|e_2)&(z_1|e_5)&(z_1|e_6)\\
(z_2|e_1)&(z_2|e_2)&(z_2|e_5)&(z_2|e_6) \end{pmatrix}
\begin{pmatrix} p\\q\\r\\s\end{pmatrix} &=
\begin{pmatrix} (e_3|b_2 + x)\\
(e_4|b_2 + x)\\
(z_1|b_2 + x)\\
(z_2|b_2 + x)
\end{pmatrix},
\\[0.3cm]
\begin{pmatrix} (e_5|e_1)&(e_5|e_2)&(e_5|e_3)&(e_5|e_4)\\
(e_6|e_1)&(e_6|e_2)&(e_6|e_3)&(e_6|e_4)\\
(z_1|e_1)&(z_1|e_2)&(z_1|e_3)&(z_1|e_4)\\
(z_2|e_1)&(z_2|e_2)&(z_2|e_3)&(z_2|e_4) \end{pmatrix}
\begin{pmatrix} p\\q\\r\\s\end{pmatrix} &=
\begin{pmatrix} (es_5|b_3 + x)\\
(e_6|b_3 + x)\\
(z_1|b_3 + x)\\
(z_2|b_3 + x)
\end{pmatrix}.\nonumber
\end{align}

\section{The Observable Matter Spectrum}\label{observable}
The basis vectors $\alpha$ and 
$\beta$
given in eq. (\ref{421}) break the $SO(10)$ symmetry
to $SU(4) \times SU(2)_L \times U(1)_R$.
Following the $\alpha$ and $\beta$ GGSO projections,
the decomposition of the spinorial $\bf{16}$ and
$\bf{\overline{16}}$ representations of $SO(10)$, 
under the $SU(4) \times SU(2)_L \times U(1)_L$ gauge group 
is given as follows:
\begin{eqnarray}
\textbf{16} &= &\left({\textbf{4}},{\textbf{2}},
0\right) + \left(\overline{{\textbf{4}}},{\textbf{1}},
-1\right) + \left(\overline{{\textbf{4}}},{\textbf{1}},
+1\right),\nonumber\\
\overline{\textbf{16}} &= &\left(\overline{{\textbf{4}}},{\textbf{2}},
0\right) + \left({\textbf{4}},{\textbf{1}},
-1\right) + \left({\textbf{4}},{\textbf{1}},
+1\right).\nonumber
\end{eqnarray}
Here to break the $SU(4) \times SU(2)_L \times U(1)_L$
gauge group to the standard model group,
we require the heavy higgs pair. This pair is given by
\begin{eqnarray}
\left(\overline{{\textbf{4}}},{\textbf{1}},
-1\right) + \left({\textbf{4}},{\textbf{1}},
-1\right).\nonumber
\end{eqnarray}
Similarly, the vectorial representation $\bf{10}$
of $SO(10)$ decomposed under the 
$SU(4) \times SU(2)_L \times U(1)_L$ 
gauge group is given as follows
\begin{eqnarray}
\textbf{10} &= &\left({\textbf{6}},{\textbf{1}},
0\right) + \left({\textbf{1}},{\textbf{2}},
-1\right) + \left({\textbf{1}},{\textbf{2}},
+1\right)\nonumber,
\end{eqnarray}
Furthermore, we take the following normalizations of the 
hypercharge and electromagnetic charge 
\begin{eqnarray}
Y &=& \frac{1}{3} (Q_1 + Q_2 + Q_3) + \frac{1}{2} (Q_4 + Q_5), \nonumber\\
Q_{em} &=& Y + \frac{1}{2} (Q_4 - Q_5). \nonumber
\end{eqnarray}
where the $Q_{i}$ charges of a state arise due to 
$\psi^{i}$ for $i =1,...,5$. The following table summaries 
the charges of the colour $SU(3)$ and
electroweak $SU(2) \times U(1)$ Cartan generators of the 
states which form the $SU(4) \times SU(2)_L \times U(1)_L$ 
matter representations:

\begin{center}
\begin{tabular}{|c|c|c|c|c|}
\hline
Representation & $\overline{\psi}^{1,2,3}$ &
$\overline{\psi}^{4,5}$ & $Y$ & $Q_{em}$ \\
\hline \hline
& ($+,+,-$)& ($+,-$)& 1/6& 2/3\\
& ($+,+,-$)& ($-,+$)& 1/6& -1/3\\
$\left( \, \textbf{4} \, , \textbf{2} , \, 0 \, \right)$ & 
$(-,-,-)$ & ($+,-$)& -1/2& 0\\
& $(-,-,-)$ & ($-,+$)& -1/2& -1\\ 
\hline
& ($+,-,-$)& $(-,-)$ & -2/3& -2/3\\
$\left( \, \overline{\textbf{4}} \, , \textbf{1} , \, -1 \, \right)$
& $(+,+,+)$ & $(-,-)$ & 0& 0\\
\hline
& ($+,-,-$)& $(+,+)$ & 1/3& 1/3\\
$\left( \, \overline{\textbf{4}} \, , \textbf{1} , \, +1 \, \right)$
& $(+,+,+)$ & $(+,+)$ & 1& 1\\ 
\hline
& ($+,-,-$)& ($+,-$)& -1/6& -2/3\\
& ($+,-,-$)& ($-,+$)& -1/6& 1/3\\
$\left( \, \overline{\textbf{4}} \, , \textbf{2} , \, 0 \, \right)$ & 
$(+,+,+)$ & ($+,-$)& 1/2& 0\\
& $(+,+,+)$ & ($-,+$)& 1/2& 1\\
\hline
& ($+,+,-$)& $(+,+)$ & 2/3& 2/3\\
$\left( \, \textbf{4} \, , \textbf{1} , \, -1 \, \right)$ & 
$(-,-,-)$ & $(+,+)$ & 0& 0\\ 
\hline
& ($+,+,-$)& $(-,-)$ & -1/3& -1/3\\
$\left( \, \textbf{4} \, , \textbf{1} , \, +1 \, \right)$
& $(-,-,-)$ & $(-,-)$ & -1& -1\\ 
\hline
\end{tabular}
\end{center}
Here $``+"$ and $``-"$,  label the contribution of an
oscillator with fermion number $F = 0$ or $F = -1$,
to the degenerate vacuum.
These states correspond to particles of the Standard Model.
More precisely we can decompose these representations under
$SU(3) \times SU(2) \times U(1)$ as
\begin{align} \label{16decomposition}
\left( \textbf{4} , \textbf{2} , 0 \right)&
= \left(\textbf{3},\textbf{2},+\frac{1}{6}\right)_{Q} + 
\left(\textbf{1},\textbf{2},-\frac{1}{2}\right)_{L}, \nonumber \\
\left( \overline{\textbf{4}} , \textbf{1} , -1 \right) &= 
\left(\overline{\textbf{3}},\textbf{1},-\frac{2}{3}\right)_{u^c}+ 
\left(\textbf{1},\textbf{1},0\right)_{\nu^c},\nonumber\\
\left( \overline{\textbf{4}} , \textbf{1} , +1 \right)&=
\left(\overline{\textbf{3}},\textbf{1},+\frac{1}{3}\right)_{d^c}+
\left(\textbf{1},\textbf{1},+1 \, \right)_{e^c}. \nonumber
\end{align}
Where $L$ is the lepton--doublet; $Q$ is the quark--doublet; 
$d^c,~u^c,~e^c$ and $\nu^c$ are the quark and lepton singlets.
Because of the $\alpha$- and $\beta$-projections, 
which projects on incomplete
$\textbf{16}$ and $\overline{\textbf{16}}$ representations, 
complete families and 
anti--families are formed by combining states from 
different sectors. 

\section{Nonviability of the 
                    $SU(4) \times SU(2) \times U(1)$ model}\label{nonv}
We now discuss why in our free fermionic construction,
the $SU(4) \times SU(2) \times U(1)$ GUT models are not viable. 
As mentioned in the previous section, 
the matter content comes from the $\bf{16}$ of $SO(10)$. 
However, with the addition of the $\alpha$ and $\beta$ basis 
vectors from eq. (\ref{421}), the $\bf{16}$ representation is 
broken by the GGSO projections that are in general given by
\begin{equation}\label{gso}
e^{i\pi v_i\cdot F_{\xi}} |S_{\xi}> = 
\delta_{{\xi}}\ C\binom {\xi} {v_i}^* |S_{\xi}>.
\end{equation}
Here $\delta_{{\xi}}=\pm1$ is a
spacetime spin statistics index
and $F_{\xi}$ is the fermion number operator.
In the SU421 models spanned by eq. (\ref{basis})
the GGSO projection coefficients $C \binom {\xi} {v_i}$ 
can take the values $\pm1; \pm i$.
Therefore, firstly considering the $\alpha$ GGSO projection,
we decompose the $\bf{16}$ into the Pati-Salam 
group representation. Moreover, using the following chirality operators
\begin{eqnarray}\label{patioperators}
X_{pqrs}^{(1)_{SO(6)}} & = &
C\binom{B^{(1)}_{pqrs}}{\alpha},\nonumber\\
X_{pqrs}^{(2)_{SO(6)}} & = &
C\binom{B^{(2)}_{pqrs}}{\alpha},\\
X_{pqrs}^{(3)_{SO(6)}} & = &
C\binom{B^{(3)}_{pqrs}}{\alpha},\nonumber
\end{eqnarray} 
we deduce that for $X_{pqrs}^{(i)_{SO(6)}} = 1$ we get the 
$Q_R\equiv(\bf{\overline{4}},\bf{1},\bf{2})$ states under 
$SU(4) \times SU(2)_L \times SU(2)_R$, whereas 
the $Q_L\equiv(\bf{4},\bf{2},\bf{1})$ states correspond to 
$X_{pqrs}^{(i)_{SO(6)}} = -1$.
Next, considering the $\beta$ GGSO projection, 
the operators 
\begin{eqnarray}\label{su4operators}
X_{pqrs}^{(1)_{421}} & = &
C\binom{B^{(1)}_{pqrs}}{\beta},\nonumber\\
X_{pqrs}^{(2)_{421}} & = &
C\binom{B^{(2)}_{pqrs}}{\beta},\\
X_{pqrs}^{(3)_{421}} & = &
C\binom{B^{(3)}_{pqrs}}{\beta}.\nonumber
\end{eqnarray} 
determine the decomposition of the $Q_L$ and $Q_R$
states under $SU(4)\times SU(2)\times U(1)$. 
Here, the product $\beta\cdot B_j^{pqrs}= -1$ with $(j=1,2,3)$,
and the modular invariance constraints, impose that
$X_{pqrs}^{(1,2,3)_{421}} = \pm \, i$. 
Therefore, this implies the states cannot be completed 
to form a family. Thus, to complete the 
$\bf{16}$ the states: $(\textbf{4} , \textbf{2} , 0)$, 
$(\overline{\textbf{4}} , \textbf{1} , -1)$ and 
$(\overline{\textbf{4}} , \textbf{1} , +1)$ under the 
$SU(4) \times SU(2)_L \times U(1)_R$ group all need to 
survive the GGSO projections, but in order for the 
$(\overline{\textbf{4}} , \textbf{1} , -1)$ and 
$(\overline{\textbf{4}} , \textbf{1} , +1)$ states to survive, 
we need $X_{pqrs}^{(1,2,3)_{421}} = \pm \, 1$, 
which is forbidden in this case by modular invariance.
To see more clearly why this is the case we consider
the decomposition of the $\bf{16}$ representation
in the combinatorial notation of ref. \cite{xmap}
\begin{eqnarray}
{\bf 16} 
& \equiv &
\left[ \binom{5}{0} + \binom{5}{2} + \binom{5}{4} \right] \label{so1016}\\
& \equiv &
\left[ \binom{3}{0} + \binom{3}{2}  \right]
\left[ \binom{2}{0} + \binom{2}{2}  \right]
~+~
\left[ \binom{3}{1} \right]
\left[ \binom{2}{1}  \right] \label{so64}\\
& \equiv &
\left[ \binom{3}{0} + \binom{3}{2}  \right]
\left[ \binom{2}{0} \right]
~+~
\left[ \binom{3}{0} + \binom{3}{2}  \right]
\left[ \binom{2}{2}  \right]
~+~
\left[ \binom{3}{1} \right]
\left[ \binom{2}{1}  \right]~~~~~
\label{su421decomposition}
\end{eqnarray}
where the combinatorial factor counts the number of periodic 
fermions in the $\vert -\rangle$ state. The second 
line in eq. (\ref{so64}) corresponds to the 
decomposition of the ${\bf16}$ under the Pati--Salam 
subgroup, whereas eq. (\ref{su421decomposition})
shows its decomposition under the SU421 subgroup. 
The key point here, as seen from 
eq. (\ref{su421decomposition}),
is the even number of fermions in the $\vert -\rangle$
vacuum of the $Q_R$ states, resulting in $\pm1$ projections
on the left--hand side of eq. (\ref{gso}), whereas the right--hand 
side is fixed by the product $\beta\cdot B_j^{pqrs}= -1$ to be $\pm i$. 
Thus, the exclusion arises because the $\beta$ projection fixes
the chirality of the vacuum of the world--sheet fermions 
${\overline\psi}^{4,5}$ that generate the $SU(2)_L\times U(1)_R$ symmetry. 
We note that the situation here is different from the 
case of the SU421 models of ref. \cite{su421}. The 
reason is that our classification method only allows for symmetric
boundary conditions for the set of internal fermions 
$\{y,\omega\vert{\overline y},{\overline\omega}\}^{1,\cdots,6}$, 
whereas the models of ref. \cite{su421} introduce 
additional freedom by allowing asymmetric boundary conditions. 
Thus, while the NAHE--based models of ref. \cite{su421} 
did not yield any model with three complete generations
they contain both the $Q_L$ and $Q_R$ states in their 
spectra, whereas vacua with only symmetric boundary conditions
with respect to the set
$\{y,\omega\vert{\overline y},{\overline\omega}\}^{1,\cdots,6}$ 
do not contain $Q_R$ states and are therefore categorically
excluded. It is of further interest to note that 
in the case of the LRS models the chirality of the
$Q_L+L_L$ and $Q_R+L_R$ is similarly affected \cite{lrs}. 
However, there it is compensated by the chirality of 
the ${\overline\eta}^j$ worldsheet fermions leading
to opposite charges under the $U(1)_j$ gauge symmetries. 
The SLM models \cite{slm} are obtained by
combining the PS and FSU5 breaking vectors.
Therefore, the SLM models produce complete ${\bf16}$
multiplets decomposed under the SLM group and 
with equal $U(1)_j$ charges. The SU421 class of 
models is the only case that is excluded
in vacua with symmetric internal boundary conditions. 

\section{Conclusion}

In this paper we discussed the classification of the 
SU421 models with symmetric internal boundary 
conditions. This continues the development
of the classification program initiated
in ref. \cite{fknr}, which led to the discovery 
of spinor--vector duality \cite{spinvecdual} 
and exophobic string vacua \cite{acfkr, cfr, SU6SU2}. 
The novel feature in the
classification of the SU421 models
compared to the PS and FSU5 vacua is the 
introduction of two basis vectors that 
break the $SO(10)$ symmetry. 
An appealing feature of the 
SU421 models is the admission of 
both the triplet--doublet as
well as the doublet--doublet 
splitting mechanism, which is 
shared only with the standard--like models. 
However, as we showed in section \ref{nonv}
these models cannot accommodate the weak  $SU(2)$ singlet
states of the Standard Model and are therefore
excluded. The next step in our classification 
program is the classification of standard--like 
models that will be reported in a future publication. 

\section{Acknowledgements}
We are grateful to John Rizos for fruitful discussions. 
The work of A.F. is partially supported by the 
UK Science and Technology Facilities Council
(STFC) under grant number ST/G00062X/1.

\medskip

\bibliographystyle{unsrt}

\end{document}